\documentstyle[aps,prd,dcolumn]{revtex}

\newcommand{\be}{\begin{equation}}
\newcommand{\ee}{\end{equation}}
\newcommand{\bea}{\begin{eqnarray}}
\newcommand{\eea}{\end{eqnarray}}
\newcommand{\nl}{\\ \nonumber}

\newcommand{\LL}{\left(\frac{\Lambda}{m}\right)}
\newcommand{\order}{{\cal O}}

\begin{document}

\title{${\cal O}(\alpha^2\Gamma,\alpha^3\Gamma)$ Binding Effects in Orthopositronium Decay}
\date{\today}
\author{Richard\ J.\ Hill \cite{email-rjh} and G.\ Peter Lepage \cite{email-gpl}}
\address{Newman Laboratory of Nuclear Studies, Cornell University\\
Ithaca, NY 14853}
\maketitle

\begin{abstract}
We present a new, simplified analysis of the low-energy 
electron-positron interaction, and use the resulting effective theory
to calculate the binding effects that contribute to the decay 
rate, $\Gamma$,
 of
orthopositronium  
through ${\cal O}(\alpha^3\ln\alpha\,\Gamma)$. We express the
total decay rate 
in terms of the annihilation rate for a free electron and
positron at threshold, which has just recently been computed to sufficient
precision. Our $\order(\alpha^2\Gamma)$ 
result corrects errors in a previous analysis.
\end{abstract}

\vspace{4mm}

%\section{Introduction}
There has been longstanding uncertainty regarding a possible 
disagreement between theoretical and experimental determinations
of the orthopositronium ground-state decay rate \cite{Sapirstein}.
A discrepancy here could
have important implications.  For example, it might indicate a 
failure of perturbative expansions in $\alpha$, 
the fine structure constant of quantum 
electrodynamics (QED); or it could signal the presence of new 
physical phenomena beyond QED.  
The decay rate is currently of particular interest
because important parts of the theoretical calculation have recently 
been completed: 
 in Ref.~\cite{Sapirstein} the annihilation rate for a free electron
and positron at threshold is calculated through $\order(m\alpha^5)$,
where $m$ is
the electron mass.  This is two
powers of~$\alpha$ beyond
leading order. 
Here we convert this decay rate for a
free electron and positron into the decay rate for a positronium atom,
the electron-positron bound state.  We also calculate new 
contributions at $\order(\alpha^3\ln\alpha\,\Gamma)$.

Traditionally, precision bound-state calculations have been formulated
within the context of quantum field theory.
A Bethe-Salpeter analysis is an example; more recently, the
nonrelativistic QED (NRQED) effective field theory approach has 
often been used\,\cite{NRQED}.  It is, however, much simpler to
recast the 
problem within the familiar framework of nonrelativistic quantum
mechanics\,\cite{Renormalizing}.  Long-range QED interactions correspond
to standard 
%instantaneous 
long-range potentials in the nonrelativistic
Hamiltonian, while short-distance effects are described by a small
number of local operators whose coefficients must be determined by 
comparing with the complete, relativistic
formulation of QED. Here we relate the short-distance
coefficients in the positronium Hamiltonian to the threshold
annihilation rate computed in~\cite{Sapirstein}, thereby obtaining the
full Hamiltonian needed for computing the decay rate.  The decay
rate
enters our formalism through the nonhermiticity of
our Hamiltonian and the complex energies that result: $\Gamma=-2{\rm Im}E$.

%The spirit of the present analysis is similar, but not identical, to
%one based on NRQED.  Here we examine the problem from the viewpoint of
%quantum mechanics, as opposed to quantum field theory.

The methods used in this paper are new to QED applications,
and have a number of desirable features.
A finite ultraviolet cutoff is built into our
Hamiltonian. This 
cutoff excludes high-momentum states that are poorly 
described by the nonrelativistic dynamics.
Unlike traditional 
approaches we do not take the cutoff to infinity at the 
end of the calculation; rather it is held fixed at a value of order
the electron's mass. Consequently,
no divergences occur, and the resulting energy eigenvalue
problem can be solved nonperturbatively using simple numerical
methods\,---\,for example, by evaluating the matrix elements of the
Hamiltonian using a finite basis set, and then solving a matrix
eigenvalue problem. 
This eliminates the need for bound-state perturbation theory.  
Another feature is that only physical inputs\,---\,on-shell
scattering amplitudes, for example\,---\,are required from full QED,
and therefore gauge and QED-regulator independence 
are explicit.  
Finally, our approach
can be adapted in a natural way to multi-electron systems such
as helium, where the lack of an exact zeroth-order solution and
the complexity of bound state perturbation theory makes a
nonperturbative solution particularly convenient.

\section{The Effective Hamiltonian}

We now proceed to the effective Hamiltonian.  We work in the 
center-of-momentum frame of the electron and positron, and consider
only states of orbital angular momentum~$0$ ($S$-states), and spin~$1$ 
(triplet- or ortho- states).   
In order to make use of the threshold results in \cite{Sapirstein},
we give the photon a small mass, $\lambda$, which is taken to~$0$ 
after the determination of the local operator
coefficients; the three final-state photons are not given a mass.

We begin with Hamiltonian
\be
\label{eq:H-approx}
H \approx \frac{p^2}{m}-\frac{p^4}{4m^3} + V + iW 
\ee
where $V$ and $W$ are hermitian.   Potential~$W$ accounts for the 
effects of three-photon decay, while $V$ accounts for all
other interactions.
To determine corrections
through~$\order(\alpha^2\Gamma)$, or~$\order(\alpha^3\Gamma)$, 
we need retain only
non-annihilation terms  
that contribute to the positronium $S$~-state binding energy 
through~$\order(\alpha^4 m)$, or~$\order(\alpha^5 m)$, respectively. 
We write $V$ as the sum of three terms,
\bea
\label{eq:V}
V(E) &=& V_0 + V_{\rm rel} + V_{\rm rad}(E),
\eea
where the Coulomb potential, $V_0$, and the leading relativistic
corrections, $V_{\rm rel}$, are given by:
\bea
\label{eq:V_0}
\langle l|V_0|k\rangle &=& \frac{-4\pi\alpha}{|l-k|^2+\lambda^2}
e^{-\frac{|l-k|^2+\lambda^2}{2\Lambda^2}} \\
\label{eq:V_rel}
\langle l|V_{\rm rel}|k\rangle &=& \left[\frac{4\pi\alpha}{|l-k|^2+\lambda^2}\
\left(\frac{2}{3m^2}|l-k|^2 -\frac{1}{2m^2}(l^2+k^2) \right)\
-\frac{4\pi\alpha}{2\Lambda^2}\
+\frac{4\pi\alpha}{4m^2}\frac{(l^2-k^2)^2}{(|l-k|^2+\lambda^2)^2}\right]
e^{-\frac{|l-k|^2+\lambda^2}{2\Lambda^2}}\nl
&&+\frac{4\pi\alpha}{2m^2}
e^{-\frac{|l-k|^2}{2\Lambda^2}}.
\eea
We have introduced Gaussian factors to suppress the potentials
at high momentum. Typically we take the ultraviolet cutoff~$\Lambda\approx m$,
although our final decay rate is (and must be) $\Lambda$-independent.
With potentials $V_0$ and $V_{\rm rel}$ in place,
our Hamiltonian correctly
reproduces the QED amplitude for
one-photon exchange and one-photon virtual annihilation 
to lowest and first order in $(v/c)^2$ (of the
electron). 
$V_{\rm rad}(E)$ accounts for one-loop radiative corrections: 
\bea
\label{eq:V_rad}
\langle l|V_{\rm rad}(E)|k\rangle &=&	\frac{8\alpha}{3\pi m^2} 
\langle l |e^{-\frac{p^2}{4\Lambda^2}}\, p^i\, 
	(p^2/m+V_0-E)\ln\left(\frac{m/4}
	{p^2/m+V_0-E}\right)\, p^i\,e^{-\frac{p^2}{4\Lambda^2}} |k\rangle \nl
&&+ \frac{14\alpha^2}{3m^2}\ln\frac{|l-k|}{m/2}
	e^{-\frac{|l-k|^2}{2\Lambda^2}}
+ \frac{\alpha^2}{m^2}\left\{
%	-\sqrt{\pi}\left[\frac{-121}{36}\frac{\Lambda}{m}-9\frac{m}{\Lambda}
%			+\frac{5}{3}\left(\frac{m}{\Lambda}\right)^3\right]
	-\frac{74}{15}+\frac{2}{3}\ln{2} + D
	\right\}e^{-\frac{|l-k|^2}{2\Lambda^2}}.
\eea
$V_{\rm rad}(E)$ gives in positronium the
%analogue of the Bethe-logarithm 
analogue of the non-recoil Lamb shift and the recoil Salpeter correction 
in hydrogen.  The final term  
is a local operator accounting for effects at high momentum; 
parameter $D$ is a counterterm which will be determined shortly.

The annihilation of the electron and positron occurs over distances of 
order $\Delta x \approx 1/m$, which are much smaller than wavelengths typical
of the electron and positron in the atom, $\lambda\approx
1/\alpha m$. Thus the annihilation potential~$W$ consists
entirely of short-distance interactions, which, to the order of
interest, can be parameterized for $S$-states as follows:
\be
\label{eq:W}
\langle l|{\rm W}|k\rangle = A^{(0)}\
\left[ (1+\alpha {A}^{(1)} +\alpha^2{A}^{(2)}
	+\alpha^3{A}^{(3)})\
%+(\frac{\beta}{2m^2}+\frac{1}{2\Lambda^2})|l-k|^2 \right]\
+(B^{(0)}+\alpha B^{(1)})\,\frac{E}{m} \right]\
e^{-\frac{|l-k|^2}{2\Lambda^2}}.
\ee
We will adjust parameters $A^{(0)}$, $A^{(1)}$, $A^{(2)}$ and  $B^{(0)}$
 so that our Hamiltonian reproduces QED results for
electron-positron annihilation into three photons.  Since 
determination of $A^{(3)}$ and of $B^{(1)}$ 
requires the as-yet unknown $\order(m\alpha^6)$
threshold annihilation rate, and the leading term in the momentum 
expansion for the $\order(m\alpha^4)$ rate, respectively, 
we simply set $A^{(3)}=B^{(1)}=0$.  Doing so 
introduces an error in the decay rate of $\order(\alpha^3\Gamma)$.

The $p^4$~operator in Eq.~(\ref{eq:H-approx}) is ill-defined at high
momenta. To 
regulate this operator, we replace it by an
energy dependent potential\,\cite{Renormalizing}:
\be 
p^4 \to m^2(E-(V+iW))^2.
\ee
Keeping only the relevant parts of this expression, our final effective
Hamiltonian is therefore:
\be \label{eq: Heff}
H(E) = H_0(E) + i \overline{W}(E)
\ee
where
\bea
H_0(E) &=&\frac{p^2}{m} + V(E) -\frac{1}{4m}\,(E-V_0)^2, \\
\overline{W}(E) &=& W +\frac{1}{2m}\,(E - V_0)(W_0+W_1).
\eea
Potentials $V$ (Eqs.(\ref{eq:V}),(\ref{eq:V_0}),(\ref{eq:V_rel}),(\ref{eq:V_rad})), 
$V_0$ (Eq.(\ref{eq:V_0})) and
$W$ (Eq.(\ref{eq:W})) are given above, and 
\bea
\langle l|W_0|k\rangle &\equiv& \langle l|W_1|k\rangle/A^{(1)} \equiv A^{(0)}e^{-\frac{|l-k|^2}{2\Lambda^2}} .
\eea

Parameter $D$ in Eq.(\ref{eq:V_rad}) is tuned to correctly reproduce  
the one-loop contribution to $e\bar{e}\to e\bar{e}$ threshold scattering,
and is found to be
\be
\label{eq:D}
D=-\sqrt{\pi}\left[\frac{-121}{36}\frac{\Lambda}{m}-9\frac{m}{\Lambda}
			+\frac{5}{3}\left(\frac{m}{\Lambda}\right)^3\right].
\ee
The hermitian Hamiltonian $H_0(E)$ is accurate through $\order(\alpha^5 m)$.
% now completely determined, 
%reproduces the $\order(\alpha^5 m)$ $^3S_1$~spectrum \cite{Yennie}.
%\bea
%\label{eq:energy}
%E(n^3S_1) &=& -\frac{m\alpha^2}{4n^2}
%	+\frac{m\alpha^4}{16n^3}\left[\frac{11}{4n}+\frac{4}{3}\right] \nl
%&&	+\frac{m\alpha^5}{8\pi n^3}
%	\left[-6\ln{\alpha}-\frac{16}{3}\ln{k_0(n)}
%		+\frac{14}{3}\left(\ln{\frac{2}{n}}+\psi(n)+\gamma
%					+\frac{1}{2n}+1\right)
%		-\frac{109}{15}+\frac{2}{3}\ln{2}\right].
%\eea
%Here $\psi(x)=\Gamma^\prime(x)/\Gamma(x)$ is the digamma function, and
%$\gamma=-\psi(1)=0.577216$ is the Euler constant.

The parameters in $W$ are  determined by considering
the imaginary part of the $S$-wave scattering amplitude for
$e\bar{e}\to 3\gamma\to e\bar{e}$,  
with electron momentum~$k$ in the center-of-momentum frame; the
optical theorem relates this
amplitude to the free-particle
annihilation rate. 
For small~$k$, the 
imaginary part of this
amplitude can be parameterized as:
\bea
\label{eq: QED}
{\cal T}(k)&=&{\cal T}_0\
\left\{\left[1 + \beta\frac{k^2}{m^2} +
\order\left(\frac{k^4}{m^4}\right)\right]
+ \alpha\left[
2\frac{m}{\lambda}+a_0 + a_1 \frac{\lambda}{m} +
\order\left(\frac{\lambda^2}{m^2}\right)\right]\right. \nl
&&\,\,\left.+\alpha^2\left[(1+2\ln{2})\frac{m^2}{\lambda^2} + 
2 a_0\frac{m}{\lambda}
-\frac{1}{3}\ln{\frac{m}{\lambda}}+ b_0 +
\order\left(\frac{\lambda}{m}\right) \right] 
+\order(\alpha^3) \right\}.
\eea
We define ${\cal T}(k)$ using nonrelativistic normalization
for the external particles\,\cite{nonrel-norm}.
Parameters ${\cal T}_0,\beta,a_0,a_1,b_0$ are determined
using QED perturbation theory~\cite{comparison}.
%; their values, including the recently calculated $b_0$,
%are given in Ref.~\cite{Sapirstein}.  

We now calculate ${\cal T}$ in our
Hamiltonian theory, and
adjust the unknown parameters to 
reproduce the QED result order by order in $\alpha$ and
$k^2$. Matching at lowest order in~$\alpha$ implies that 
\be
A^{(0)} = {\cal T}_0, \quad\quad\quad
B^{(0)} = \beta+\frac{m^2}{\Lambda^2}.
\ee
$A^{(1)}$ and $A^{(2)}$ are determined by matching 
the $\order(\alpha)$ and $\order(\alpha^2)$ contributions in ${\cal T}(0)$,
respectively; we obtain
\bea
A^{(1)} &=& a_0 +
\frac{1}{\sqrt{\pi}}\left[\frac{4}{3}\LL+3\LL^{-1}\right]
 + \left(a_1 -\frac{7}{12} +\LL^{-2}\right)\frac{\lambda}{m}\
+ \order\left(\frac{\lambda^2}{m^2}\right), \\
A^{(2)} &=& b_0-2a_1
+\frac{1}{\sqrt{\pi}}\left[\frac{4}{3}\LL+3\LL^{-1}
			\right]a_0
	+\frac{1}{3}\ln\frac{\Lambda}{m}
	+\frac{1}{\pi\sqrt{\pi}}\left\{\left[ -\frac{44\sqrt{6}}{81}
				\left(\gamma-\ln{\frac{2\Lambda^2}{3m^2}}
				-2\right)
		\right]\LL^3\right.\nl
&&\left.	+\left[
		\frac{7}{3}\ln{\frac{\Lambda}{m}}
		+\frac{56\sqrt{6}}{27}\left(\gamma-\ln{\frac{2\Lambda^2}{3m^2}}
						-\frac{2}{7}\right)
		-\frac{37}{15} 
		+\frac{1}{3}\ln{2}-\frac{7}{6}\gamma
	\right]
		\LL\right\}
	+\left(\frac{83}{24\pi}-\frac{11\sqrt{3}}{12\pi}+\frac{11}{48}\right)
		\LL^2\nl
&&	+\left(\frac{25}{2\pi}
		-\frac{4\sqrt{3}}{3\pi}+\frac{17}{18}
		-\frac{5}{6}\ln{2}-\frac{1}{3}\gamma
		+\frac{2}{\sqrt{\pi}}\kappa\right)
+\left(\frac{49}{6\pi}-\frac{3\sqrt{3}}{2\pi}-\frac{1}{4}\right)\LL^{-2}
+\order\left(\frac{\lambda}{m}\right).
\eea
Here $\gamma=-\psi(1)=0.577216$ is the Euler constant and 
$\kappa\equiv\int dx\ \ln{x^{-1}}\exp{(-x^2)}{\rm erf}(x)^2 =0.051428$. 
Having determined all 
the necessary parameters, we can now safely set $\lambda = 0$.

\section{The Decay Rate}
Now that our Hamiltonian is completely specified
we finally solve for the decay rate, given by the imaginary part
of the ground state energy eigenvalue. 
Note that due to the presence of the cutoff, no divergences 
occur when calculating matrix elements, and no intricate limiting 
procedures are necessary to solve the eigenvalue
problem\,---\,renormalization is automatic.
%We could simply evaluate
%matrix elements of the full Hamiltonian $H=H_0+i\overline{W}$
%(Eq.~(\ref{eq: Heff})) using a convenient  
%basis set, and then solve the resulting nonhermitian matrix eigenvalue 
%problem. The imaginary part of the
%ground-state eigenvalue gives the decay rate.  
To avoid dealing with nonhermitian matrices we choose to
work only to first order in the annihilation potential~$\overline{W}$;
higher-order terms are suppressed 
by several powers of $\alpha$ beyond the precision of interest.

We first solve the eigenvalue problem for~$H_0$,
\be
H_0(E_0)\, |\psi_0\rangle = E_0 \,|\psi_0\rangle,
\ee
to obtain the ground-state energy and wavefunction.
The energy dependence of $H_0$ is easily handled by iterating the
eigenvalue equation, starting with an approximate energy in~$H_0$;
the answer converges to adequate precision after only a few iterations.
The eigenfunctions for our energy dependent Hamiltonian must be normalized
so that\,\cite{Lepage-thesis}
\be
\left.\langle \psi_0|1-\frac{\partial H_0}{\partial E}
|\psi_0\rangle\right|_{E=E_0} = 1.
\ee
Then the decay rate, to first order in $\overline W$, is 
\bea
\Gamma&=&-2\langle \psi_0 | \overline{W}(E_0) | \psi_0 \rangle\\
\label{eq: gamma}
&=& \Gamma_0\left\{ 
        \left[1+\alpha A^{(1)} + \alpha^2 A^{(2)} + \alpha^3 A^{(3)} 
	+\left(\frac{1}{2}\left(1+\alpha A^{(1)}\right)
		+B^{(0)}+\alpha B^{(1)}\right)
		\frac{E_0}{m} \right]\,\alpha^2 M_1
%	\langle \delta^3_\Lambda(r)\rangle 
%	\right.\\
%&&\left.	
	+\frac{1}{2}(1 + \alpha A^{(1)})\,\alpha^2 M_2
%	\frac{\langle\alpha v_\Lambda(r)\delta^3_\Lambda(r)\rangle}{2m} 
	\right\}
\eea
where $\langle {\cal O} \rangle \equiv \langle \psi_0 | {\cal O} |
\psi_0\rangle$,
$\langle \delta^3_\Lambda(r) \rangle \equiv (m^3\alpha^5/8\pi)M_1$,
$\langle \alpha v_\Lambda(r)\delta^3_\Lambda(r) \rangle \equiv
(m^4\alpha^5/8\pi)M_2$ and
$\Gamma_0$ is the lowest order $1S$ decay rate.
The cutoff operators $v_\Lambda(r)$ and
$\delta^3_\Lambda(r)=-\nabla^2v_\Lambda(r)/4\pi$ are defined by 
Fourier transform:
\be
\frac{4\pi}{q^2}\, e^{-\frac{q^2}{2\Lambda^2}}\quad
{\longrightarrow}\quad v_\Lambda(r) \equiv \frac{1}{r}\,{\rm
erf}\left(\frac{\Lambda 
r}{\sqrt{2}}\right) . 
\ee

The matrix elements $M_1$, $M_2$,
in Eq.\,(\ref{eq: gamma}) can be evaluated for any 
$S$-state of $H_0$ and the corresponding decay rate computed using
this equation. We used  bases consisting of~30 to~60 gaussians, with
varying widths, to diagonalize~$H_0$.  The numerical eigenvalues 
accurately reproduce the $^3S_1$ spectrum through 
$\order(\alpha^5 m)$~\cite{Yennie}.
Our ground-state results, for
several values 
of~$\Lambda$, are shown in
Table~\ref{table: Lambda}; there we introduce the dimensionless parameter
$X_\Gamma$, defined for any $S$-state 
by:
\be
\label{eq:Gamma_def}
\Gamma(nS) \equiv \frac{\Gamma_0}{n^3}\left[1+\alpha a_0 
	+\alpha^2\left(\frac{1}{3}(1+\alpha a_0)\ln{\alpha}
	+b_0-2a_1-\frac{\beta}{4n^2}
	-\frac{3}{2\pi}\alpha\ln^2\alpha
	+X_\Gamma(nS)\right)\right] .
\ee
%where $\Gamma_0$ is the lowest order 1$S$ decay rate.
%\be
%\langle \delta^3_\Lambda(r) \rangle \equiv \frac{m^3\alpha^5}{8\pi}M_1,
%\quad\quad 
%\langle \alpha v_\Lambda(r)\delta^3_\Lambda(r) \rangle \equiv
%\frac{m^4\alpha^5}{8\pi}M_2. \nl 
%\ee
This definition anticipates the leading 
$\alpha^2\ln\alpha$~\cite{Caswell-Lepage} and 
$\alpha^3\ln^2\alpha$~\cite{Karshenboim} 
contributions, which are correctly reproduced by our numerical analysis.
The final results for $X_\Gamma$ are almost independent of
$\Lambda$, while the changes in the matrix elements from one~$\Lambda$
to the next are two orders of
magnitude larger than~$X_\Gamma$ itself.
Renormalization theory
guarantees that $\Lambda$ dependence due to
the matrix elements cancels, in the final answer, against
$\Lambda$ dependence due to coefficients~$A^{(1)}$, $A^{(2)}$ and $B^{(0)}$
in the annihilation potential~$W$.
The residual $\Lambda$ dependence
in~$X_\Gamma$ is due to our approximations in potentials~$V$ and~$W$,
the dominant effect being
at~$\order(\alpha^3\Gamma)$ where we have left out the contributions 
from~$A^{(3)}$ and $B^{(1)}$.
%\be
%\delta\Gamma = -2{\cal T}_0\,\alpha^3 A^{(3)}
%	\langle \delta^3_\Lambda(r) \rangle 
%	= \frac{\Gamma_0}{n^3}(\alpha^3 A^{(3)}+\order(\alpha^4) ).
%\ee

Our calculation is nonperturbative in potential~$V$ and so
automatically includes order~$\alpha^3\Gamma$
(and higher-order) corrections to the decay rate. 
To facilitate comparison with other calculations, we suppressed these
higher-order effects by extrapolating our analysis to~$\alpha=0$. 
Our results for $\Lambda=m$ and a range of $\alpha$'s are shown in 
Table~\ref{table: alpha}. Upon extrapolation to $\alpha=0$, we obtain:
\be
\left.X_\Gamma(n)\right|_{\alpha\to0} =
\left\{
\begin{array}{cl}
0.913, & n=1S \\
2.588, & n=2S \\
2.936, & n=3S \\
\end{array}
\right.
\ee
%\bea
%\left.X_\Gamma(1S)\right|_{\alpha\to0} &=& 0.913 -0.2500\beta\nl
%\left.X_\Gamma(2S)\right|_{\alpha\to0} &=& 2.588 -0.0625\beta\nl
%\left.X_\Gamma(3S)\right|_{\alpha\to0} &=& 2.934 -0.0278\beta
%\eea
where all results are accurate to within~$1$ in the last digit.
The 1$S$ result agrees well with the analytic result in~\cite{Sapirstein}.
%The exact $1/n^2$ dependence of the coefficient of $\beta$ agrees 
%with the prediction in~\cite{Labelle-PRL}.

Since the analysis would be complete through $\order(\alpha^3\Gamma)$ with
the inclusion of the operators parameterized by~$A^{(3)}$ and~$B^{(1)}$,
 and since
neither of these operators generates factors of $\ln\alpha$, the complete 
contribution in the decay rate of  $\order(\alpha^3\ln\alpha\,\Gamma)$
is already present, along with that of  $\order(\alpha^3\ln^2\alpha\,\Gamma)$.
By examining the $\alpha$ dependence of $X_\Gamma$, we find:
\be
\label{eq:Xresult}
X_\Gamma(1S) = 0.913-0.665\, \alpha\ln\alpha +\order(\alpha) .
\ee
The coefficient of $\alpha\ln\alpha$ is independent of the cutoff $\Lambda$.
The non-logarithmic term of $\order(\alpha^3\Gamma)$ 
is cutoff-dependent, since we have neglected
contributions from the cutoff-dependent~$A^{(3)}$ and $B^{(1)}$.
Using the values for~$a_0$ and $\Gamma_0$
%, and the theoretical 
%prediction through $\order(\alpha^3\ln^2{\alpha}\Gamma)$ 
in Ref.~\cite{Sapirstein}, 
the $\order(\alpha^3\ln{\alpha}\,\Gamma)$ 
contribution amounts to
$2.4\times 10^{-5}\mu s^{-1}$.  Together with a small contribution
from five-photon decay of $0.73\times 10^{-5}\mu s^{-1}$~\cite{five-photon},
this brings the current 
theoretical prediction for the decay rate to 
$\Gamma=7.039967(10)\mu s^{-1}$.
%~\cite{comparison}
%~\cite{Kniehl}

After completing our calculation, we learned of an independent
analysis of $\order(\alpha^3\ln\alpha\Gamma)$ contributions to
positronium decay by Kniehl and Penin~\cite{Kniehl}.  Our result,
$-0.665$ for the coefficient independent of $a_0$, disagrees 
with their result, $-(4/5+8\ln{2}/3)/\pi=-0.8430$.   
%In this 
%situation, it is desirable to have a cross-check with other 
%calculations.  Available in the literature are calculations
%at $\order(\alpha^3\ln{\alpha})$ in the muonium 
%hyperfine splitting (HFS) which involve the same operators,
%with different coefficients, as those contributing at this 
%order in the decay rate.    
To verify our analysis, we compared it with published results
on $\order(\alpha^3\ln\alpha)$ contributions to muonium 
hyperfine splitting (HFS).  The HFS results involve the same
operators, with different coefficients, as those contributing
at the same relative order in the decay rate.
Terms in the orthopositronium
decay rate not involving $a_0$ arise from second order perturbations
of $V_{\rm rad}$ with the leading decay operator: 
\be
\label{eq:Gamma_op}
\frac{\delta\Gamma}{\Gamma_0} = 2\frac{\langle V_{\rm rad}\tilde{G}\delta^3(r)\rangle}{\langle\delta^3(r)\rangle} +\langle\frac{\partial V_{\rm rad}}{\partial E}\rangle,
\ee
where $\tilde{G}$ is the Coulomb Green's function with the ground state
pole removed, and expectation values are taken between unperturbed 
Coulomb eigenfunctions.  
$V_{\rm rad}$ can be expressed as 
\be
\label{eq:V_rad_op}
V_{\rm rad} = 
\frac{2}{3}\alpha^2{\cal O}_1 
+ \frac{7}{6}\alpha^2{\cal O}_2 
+ \left(\frac{1}{6}\ln{2}-\frac{37}{30}\right)\alpha^2{\cal O}_3,
\ee
where
\be
\label{eq:O_i}
{\cal O}_1 = \frac{1}{\pi\alpha m_r^2}p^i\left(\frac{p^2}{2m_r}-\frac{\alpha}{r}-E\right)\ln\frac{m_r/2}{\frac{p^2}{2m_r}-\frac{\alpha}{r}-E}\,p^i,
\quad\langle l|{\cal O}_2|k\rangle = \frac{1}{m_r^2}\ln\frac{|l-k|}{m_r},
\quad\langle l|{\cal O}_3|k\rangle = \frac{1}{m_r^2}.
\ee
The reduced mass $m_r$ equals $m/2$ in positronium.  
%The logarithmic parts of the matrix elements for the 
%second order perturbations, which are independent of the cutoff, 
%are given by:
The logarithmic parts of the matrix elements for the second-order 
perturbations can be inferred, for the most part, directly from the
HFS papers \cite{Kinoshita-PRD}\cite{Kinoshita-talk}:
\be
\label{eq:second-order}
\left(2\frac{\langle {\cal O}_i\tilde{G}\delta^3(r)\rangle}{\langle \delta^3(r)\rangle} +  \langle\frac{\partial {\cal O}_i}{\partial E}\rangle\right) 
\to \frac{\alpha}{\pi} \times 
\left\{
\begin{array}{cl}
-4\ln^2\alpha-8(-\ln 2+3/4)\ln\alpha & ,\ i=1 \\
\ln^2\alpha+(2\ln{2}-1)\ln\alpha & ,\ i=2 \\
2\ln\alpha & ,\ i=3 \\
\end{array}
\right.
\ee
The only exception is the coefficient $(2\ln{2}-1)$ of $\ln\alpha$ for 
${\cal O}_2$.  A partial analysis in Ref.~\cite{Kinoshita-talk} gives
$(2\ln{2}+2)$ in place of $(2\ln{2}-1)$.  We have calculated the full
contribution analytically, and find the result shown in 
Eq.~(\ref{eq:second-order})~\cite{muonium}.
We have also verified the results in Eq.(\ref{eq:second-order}) by
direct numerical evaluation, using our Gaussian basis set.  
Non-logarithmic terms of $\order(\alpha^0,\alpha^1)$ are 
cutoff dependent, but the logarithmic terms were cutoff independent,
as expected.  
%These expressions can be inferred
%from HFS papers \cite{Kinoshita-PRD}\cite{Kinoshita-talk}, with the 
%exception of the single logarithm coefficient, 
%$(2\ln 2 - 1)$, for ${\cal O}_2$.  
%In Ref.~\cite{Kinoshita-talk}, which would give $(2\ln{2}+2)$ in place of $(2\ln{2}-1)$, it appears
%that this value was deduced by approximating the constant part of the
%recoil Salpeter correction (there denoted by $C_S$) as the result of
%a delta-function potential.  We have recalculated this term, and find the
%above result~\cite{muonium}.  
%Eqs.~(\ref{eq:Gamma_op}), (\ref{eq:V_rad_op}), (\ref{eq:second-order})
%give an analytic expression for the decay rate.  
Combining our analytic results with those in Ref.\cite{Sapirstein},
our Eq.~(\ref{eq:Xresult}) becomes:
%If we assume the analytic form
%for the $\order(\alpha^2\Gamma)$ contribution found in Ref.\cite{Sapirstein},
%our Eq.~(\ref{eq:Xresult}) would be:
\bea
X_\Gamma(1S) &=& \frac{11}{8}-\frac{2}{3}\ln{2}+ 
\left(8\ln{2}-\frac{229}{30}\right)\frac{\alpha}{\pi}\ln{\alpha} \nl
&=& 0.9129 -0.6647 \alpha\ln{\alpha},
\eea
in complete agreement with our numerical analysis. 

We thank the authors of~\cite{Sapirstein} for sharing their results
with us before they were published. We also thank Patrick Labelle for
several discussions. This work was supported by a grant 
from the National Science Foundation.

\begin{table}
\begin{center}
\begin{tabular}{crrr} 
%\toprule
 
\hline\hline
{$\Lambda/m$} &  $M_1$ & \mbox{$M_2$} & \mbox{$X_\Gamma$} \\ 
%\colrule
 
\hline
%$0.25$ & $17839.5823$ & $18.4265$ & $0.5846$ \\
$0.5$ & $18249.0812$ & $37.6279$ & $0.8518$ \\
$1.0$ & $18405.3892$ & $75.8182$ & $0.8867$ \\
$1.5$ & $18410.4321$ & $113.7030$ & $0.8323$ \\
$2.0$ & $18375.7156$ & $151.2649$ & $0.5852$ \\
\hline\hline
\end{tabular}
\end{center}
\caption{Matrix elements and corrections to the ground state
decay rate evaluated at
$\alpha^{-1}=137.03599976$ as a function of cutoff $\Lambda$.
To determine $X_\Gamma$, we set $a_0=a_1=b_0=0$.
} 
\label{table: Lambda}
\end{table}

\begin{table}
\begin{center}
\begin{tabular}{lccc}
%\toprule
\hline\hline
\mbox{$\alpha$} &
$X_\Gamma(1S)$ & $X_\Gamma(2S)$ & $X_\Gamma(3S)$\\ 
%\colrule
 
\hline
0.08 
& $0.5362\quad$
& $2.1012\quad$ 
& $2.4437\quad$\\
0.04 
& $0.7299\quad$
& $2.3607\quad$
& $2.7094\quad$\\
0.02 
& $0.8282\quad$
& $2.4870\quad$
& $2.8373\quad$\\
0.01 
& $0.8749\quad$
& $2.5449\quad$
& $2.8949\quad$\\
0.005
& $0.8962\quad$
& $2.5703\quad$
& $2.9198\quad$\\
0.0025
& $0.9057\quad$
& $2.5812\quad$
& $2.9305\quad$\\
\hline
\mbox{$\to0$} &$ 0.913 \quad $ 
& $2.588 \quad$ & $2.936 \quad$ \\
\hline\hline
\end{tabular}
\end{center}
\caption{Corrections to decay rate at $\Lambda=m$ 
as a function of $\alpha$.
Here $A^{(1)}=a_0+2.444821528$ \\
and 
$A^{(2)}=b_0 - 2 a_1+2.444822 a_0 + 6.179923$.
}
\label{table: alpha}
\end{table}


\begin{thebibliography}{99}
\bibitem{comparison} The values of these parameters 
can be found in Ref.~\cite{Sapirstein}.
By comparing our Eq.~(\ref{eq: QED}) to
Eqs.~(8), (10) and (12) of Ref.~\cite{Sapirstein}, the correspondence 
between our parameters and theirs must be: 
$a_0+\lambda a_1 = A(\lambda)/\pi = (-10.286606(10)+15.39\lambda)/\pi$,
$b_0-2a_1=B_2/\pi^2=22.38(26)/\pi^2$ and  
$\beta=-X/12=-(19\pi^2-132)/12(\pi^2-9)$.  
Five-photon decay contributes an additional
$0.19(1)/\pi^2$ to $b_0$; see Ref.~\cite{five-photon}.
\bibitem{nonrel-norm}
The amplitude calculated using the usual relativistically
invariant normalization should be multiplied by a factor 
$1/\sqrt{2{E}}$ for each external particle, 
with ${E}=\sqrt{m^2+k^2}$, to obtain this nonrelativistic 
normalization.
\bibitem{muonium} The additional HFS contribution with $(2\ln{2}-1)$
in place of $(2\ln{2}+2)$ amounts to $0.18\,{\rm kHz}$ in muonium.  Taken 
alone this would affect the theoretical prediction for the HFS,
and consequently the determination of the muon mass, at 
the $1.5\sigma$ level.  
%Part of this difference may be accounted for
%by the ``constant'' term, $0.11\pm 0.06\,{\rm kHz}$
%in Ref.~\cite{Kinoshita-talk}.
We are preparing a paper to clarify this situation,
giving complete results on $\order(\alpha^3\ln\alpha)$
corrections for muoniun and  positronium HFS.
\end{thebibliography}
\end{document}